\documentclass[aps,prl,amsmath,twocolumn]{revtex4}
\usepackage{bm}
\usepackage{graphicx}
\begin{document}
\title{Nonlocal Andreev reflection at high transmissions}
\author{Mikhail S. Kalenkov and Andrei D. Zaikin}
\affiliation{Forschungszentrum Karlsruhe, Institut f\"ur Nanotechnologie,
76021, Karlsruhe, Germany}
\affiliation{I.E. Tamm Department of Theoretical Physics, P.N.
Lebedev Physics Institute, 119991 Moscow, Russia}

\begin{abstract}
We analyze non-local effects in electron transport across
three-terminal normal-superconducting-normal (NSN) structures.
Subgap electrons entering S-electrode from one N-metal may form
Cooper pairs with their counterparts penetrating from another
N-metal. This phenomenon of crossed Andreev reflection -- combined
with normal scattering at SN interfaces -- yields two different
contributions to non-local conductance which we evaluate
non-perturbatively at arbitrary interface transmissions. Both
these contributions reach their maximum values at fully
transmitting interfaces and demonstrate interesting features which
can be tested in future experiments.
\end{abstract}


\maketitle

At sufficiently low temperatures Andreev reflection (AR)
\cite{And} dominates charge transfer through an interface between
a normal metal and a superconductor (NS): An electron propagating
from the normal metal with energy below the superconducting gap
$\Delta$ enters the superconductor at a length of order of
the superconducting coherence length $\xi$, forms a Cooper pair
together with another electron, while a hole goes back into the
normal metal. As a result, the net charge $2e$ is transferred through
the NS interface which acquires non-zero subgap conductance
\cite{BTK}.

In hybrid NSN structures with two N-terminals, electrons may
penetrate into a superconductor through both NS interfaces.
Provided the superconductor size (distance between two NS
interfaces) $L$ strongly exceeds $\xi$, AR processes at these
interfaces are independent. If, however, the distance $L$ is
smaller than or comparable with $\xi$, two additional non-local
processes come into play (see Fig. 1). Firstly, an electron
with subgap energy propagating from one N-metal can penetrate through the
superconductor into another N-electrode with the probability $\sim
\exp (-L/\xi )$. Secondly, an electron penetrating into the
superconductor from the first N-terminal may form a Cooper pair by
``picking up'' another electron from the second N-terminal. In
this case a hole will go into the second (not the first!) N-metal
and, hence, AR turns into a non-local effect. The probability of
this process -- usually called crossed Andreev reflection (CAR)
\cite{BF,GF} -- also decays as $\sim \exp (-L/\xi )$ and, in
combination with direct electron transfer between normal
electrodes, determines non-local conductance in hybrid
multi-terminal structures which can be directly measured in
experiment.

\begin{figure}
\includegraphics[width=6.5cm]{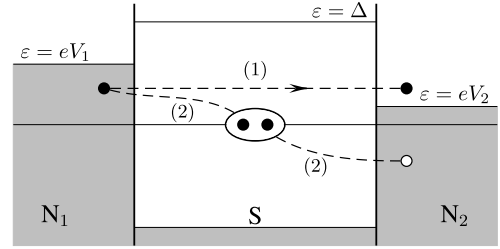}
\caption{Two elementary processes contributing to non-local
conductance of an NSN device: (1) direct electron transfer and (2)
crossed Andreev reflection.}
\end{figure}

CAR has recently become a subject of intensive investigations both
in experiment \cite{Beckmann,Teun,Venkat} and in theory
\cite{FFH,Fabio,Melin,Belzig,BG} (see also further references
therein). Although a non-local conductance was observed in all
these experiments, an unambiguous and detailed interpretation of
the existing experimental data still remains a challenge, to a
certain extent because in addition to the above processes a number
of other physical effects may considerably influence the
observations. Among such effects we mention, e.g, charge imbalance
(relevant close to the superconducting critical temperature
\cite{Beckmann,Venkat}) as well as zero-bias anomalies in the
Andreev conductance due to both disorder-enhanced interference of
electrons \cite{VZK,HN,Z} and Coulomb effects \cite{Z,HHK,GZ06}.
CAR is also sensitive to magnetic properties of normal electrodes.
Although theoretical investigation of the above physical effects
is certainly of interest and may help to account for some
experimental observations, we believe that, beforehand, it is
important to reach quantitative understanding of CAR in simpler
situations when (at least some of) the above effects can be
disregarded.

As in most cases metallic interfaces are not fully transparent, AR
is usually combined with normal electron scattering at such
interfaces. The relative ``weights'' of these two processes are
determined by interface transmission. In the case of multi-terminal
hybrid structures normal reflection, tunneling, local AR and CAR
combine in a complicated and non-trivial manner. For instance, it
was demonstrated \cite{FFH,Fabio} that in the lowest order in the
interface barrier transmission and at $T=0$ CAR contribution to
cross-terminal conductance is exactly cancelled by that from
elastic electron cotunneling \cite{FN}, while no such cancellation
is expected in higher orders in the transmission \cite{Melin}.
However, complete theory of non-local phenomena in question which
would fully describe an interplay between all scattering processes
to all orders in the interface transmissions and set the maximum
scale of the effect remains unavailable. Such a theory requires
non-perturbative methods and is the main subject of the present
work.

\begin{figure}
\includegraphics[width=6.cm]{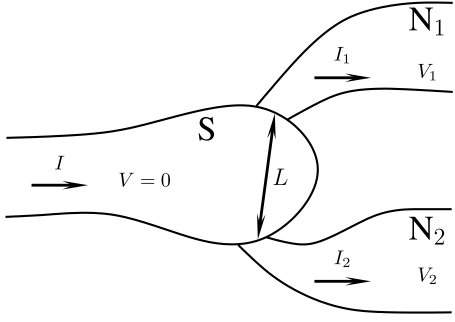}
\caption{Schematics of our NSN device.}
\end{figure}

{\it The model and formalism}. Consider three-terminal NSN
structure depicted in Fig. 2. We will assume that all three
metallic electrodes are non-magnetic and ballistic, i.e. the
electron elastic mean free path is large. Transmissions $D_1$ and
$D_2$ of two SN interfaces (with cross-sections ${\cal A}_1$ and
${\cal A}_2$) may take any value from zero to one. The distance
between the two interfaces $L$ as well as other geometric
parameters are assumed to be much larger than $\sqrt{{\cal
A}_{1,2}}$, i.e. effectively both contacts are metallic
constrictions. In this case the voltage drops only across SN
interfaces and not inside large metallic electrodes. Hence,
nonequilibrium (e.g. charge imbalance) effects related to the
electric field penetration into the S-electrode can be neglected.
In what follows we will also ignore Coulomb effects
\cite{Z,HHK,GZ06}.

For convenience, we will set the electric potential
of the S-electrode equal to zero, $V=0$. In the presence of bias voltages
$V_1$ and $V_2$ applied to two normal electrodes (see Fig. 2) the currents
$I_1$ and $I_2$ will flow through SN$_1$ and SN$_2$ interfaces. These currents
can be evaluated with the aid of the quasiclassical formalism of
nonequilibrium Green-Eilenberger-Keldysh functions $\hat g^{R,A,K}$
\cite{BWBSZ}. In the ballistic limit the corresponding equations
take the form
\begin{gather}
\begin{split}
\left[
\varepsilon \hat\tau_3+
eV(\bm{r},t)-
\hat\Delta(\bm{r},t),
\hat g^{R,A,K} (\bm{p}_F, \varepsilon, \bm{r},t)
\right]
+\\+
i\bm{v}_F \nabla \hat g^{R,A,K} (\bm{p}_F, \varepsilon, \bm{r},t) =0,
\end{split}
\label{Eil}
\end{gather}
where $[\hat a, \hat b]= \hat a\hat b - \hat b \hat
a$, $\varepsilon$ is the quasiparticle energy, $\bm{p}_F=m\bm{v}_F$ is the
electron Fermi momentum vector and $\hat\tau_3$ is the Pauli matrix.
The functions  $\hat g^{R,A,K}$ also obey the normalization conditions
$(\hat g^R)^2=(\hat g^A)^2=1$ and $\hat g^R \hat g^K + \hat g^K \hat g^A =0$.
Here and below the product of matrices is defined as time convolution.

The matrices  $\hat g$ and $\hat\Delta$ have the standard form
\begin{equation}
        \hat g^{R,A,K} =
        \begin{pmatrix}
                g^{R,A,K} & f^{R,A,K} \\
                \tilde f^{R,A,K} & \tilde g^{R,A,K} \\
        \end{pmatrix}, \quad
        \hat\Delta=
        \begin{pmatrix}
                0 & \Delta \\
                -\Delta^* & 0 \\
        \end{pmatrix},
\end{equation}
where $\Delta$ is the BCS order parameter. The current density is
related to the Keldysh function $\hat g^K$ as
\begin{equation}
\bm{j}(\bm{r}, t)= -\dfrac{e N_0}{4} \int d \varepsilon
\left< \bm{v}_F \mathrm{Sp} [\hat \tau_3 \hat g^K(\bm{p}_F,
\varepsilon, \bm{r},t)] \right>,
\label{current}
\end{equation}
where $N_0=mp_F/2\pi^2$ is the density of state at the Fermi level and
angular brackets $\left< ... \right>$ denote averaging over the Fermi momentum
directions.

The above equations should be supplemented by appropriate boundary
conditions. In order to match quasiclassical Green functions
at the N- and S-sides of SN$_1$ interface (respectively $\check
g_{N_1}$ and $\check g_S$) we will make use of Zaitsev boundary
conditions \cite{Zaitsev} for matrices $\check g= \left( \begin{smallmatrix}
\hat g^R & \hat g^K \\ 0 & \hat g^A \end{smallmatrix}\right)$:
\begin{gather}
\check g^a = \check g_{N_1}^+ - \check g_{N_1}^- = \check g_S^+ - \check g_S^-,
\label{cont}
\\
\check g^a [ R_1 (\check g^+)^2 + (\check g^-)^2]=D_1 \check g^-
\check g^+,
\label{bcond}
\end{gather}
where $\check g^\pm = \check g_{N_1}^+ + \check g_{N_1}^- \pm
\check g_S^+ \pm \check g_S^-$, $\check g_S^\pm =\check g_S(\pm
p_x)$ (see Fig. 3a), $R_1(p_{x_1})\equiv 1-D_1(p_{x_1})$,
$p_{x_1}$ is the component of $\bm{p}_F$ normal to the SN$_1$
interface. Green functions at SN$_2$ interface are matched
analogously. Deep inside metallic electrodes S, N$_1$ and N$_2$
the Green functions should approach their equilibrium values $\hat
g^{R,A}=\pm (\varepsilon\hat\tau_3-\hat\Delta) /\Omega^{R,A}$ in a
superconductor and $\hat g^{R,A}=\pm \hat\tau_3$ in normal metals,
$\Omega^{R,A} =\sqrt{(\varepsilon \pm i\delta)^2-\Delta^2}$. For
the Keldysh functions far from interfaces we have $\hat g^K = \hat g^R 
\left(\begin{smallmatrix}
h_+ & 0 \\ 0 & h_-
\end{smallmatrix}\right) - 
\left(\begin{smallmatrix}
h_+ & 0 \\ 0 & h_-
\end{smallmatrix}\right)
\hat g^A$, where  $h_{\pm}=\tanh [(\epsilon \pm eV)/2T]$.
Voltage in above expression equals to $V=0$, $V_1$ and $V_2$ respectively 
in S, N$_1$ and N$_2$ electrodes. The parameter $\Delta$ is chosen to be real.

\begin{figure}
\includegraphics[width=7.cm]{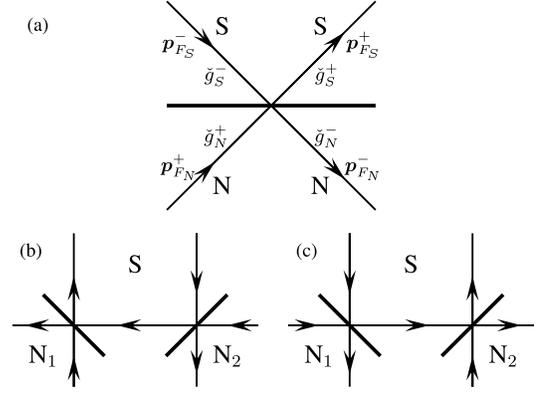}
\caption{Quasiclassical trajectories contributing to local (a) and non-local
  (b and c) currents.}
\end{figure}

{\it Relevant trajectories}. Electron trajectories which
contribute to the current $I_1$ through SN$_1$interface are shown
in Fig. 3. Trajectories presented in Fig. 3a do not enter the
terminal N$_2$ and yield the standard BTK contribution \cite{BTK} to $I_1$.
In addition there exist trajectories (Fig. 3b,c) involving all
three electrodes. They fully account for all scattering processes
-- both normal and AR -- to all orders in the interface
transmissions and determine non-local conductance of our NSN
device. As follows from Fig. 3b,c for each direction of $p_x$ one
can distinguish four different contributions to non-local
conductance corresponding to different trajectory combinations.

Note that applicability of the above quasiclassical formalism with
boundary conditions (\ref{bcond}) to hybrid structures with two
(or more) barriers is, in general, a non-trivial issue \cite{GZ02}
which requires a comment. Electrons scattered at different
barriers may interfere and form bound states (resonances) which
cannot be correctly described within a formalism employing Zaitsev
boundary conditions \cite{Zaitsev}. In our geometry, however, any
relevant trajectory reaches each interface only once whereas the
probability of multiple reflections at both interfaces is small in
the parameter ${\cal A}_1 {\cal A}_2/L^4 \ll 1$. Hence, resonances
formed by multiply reflected electron waves can be neglected, and
our formalism remains adequate for the problem in
question.

\textit{Quasiclassical Green functions}. The above equations can
be conveniently solved introducing parameterization of the matrix
Green functions $\hat g^{R,A,K}$ by four Riccati amplitudes and
two ``distribution functions'' \cite{Eschrig}. This
parameterization allows to transform Eq. (\ref{Eil}) to a set of
decoupled equations. It is also important that non-linear Zaitsev
boundary conditions (\ref{cont}), (\ref{bcond}) can be rewritten
in terms of Riccati amplitudes and ``distribution functions'' in a
rather simple form \cite{Eschrig}. Integration of the resulting
equations along the trajectories shown in Fig. 3 is
straightforward. Finally we arrive at the following expression for
the Keldysh Green function $g^K_{N_1}$ at SN$_1$ interface (on the
N-metal side)
\begin{equation}
g_{N_1}^K=g_{1,a}^K(V_1)+g_{1,b+c}^K(V_1)+g_{12,b+c}^K(V_2).
\label{gN1}
\end{equation}
Here $g_{1,a}^K(V_1)$ comes from the trajectories of Fig. 3a
responsible for the BTK current at SN$_1$ interface, while two
other terms come from the trajectories of Fig. 3b,c which also
involve N$_2$-electrode. The term $g_{1,b+c}^K(V_1)$ yields a
correction to the BTK term which will be discussed later. The last
contribution $g_{12,b+c}^K(V_2)$ accounts for non-local
conductance of our device. For positive $p_{x_1}>0$ we have
\begin{multline}
g^K_{12,b+c}(V_2)=2 D_1 D_2
\dfrac{1-\tanh^2iL\Omega/v_F}{P(R_1,R_2)}
\\\times
\Biggl(\theta_c R_1 R_2 |a|^4\tanh \dfrac{\varepsilon + eV_2}{2T}+
\theta_b R_2 |a|^2 \tanh \dfrac{\varepsilon - eV_2}{2T}
\\+
\theta_c R_1 |a|^2 \tanh \dfrac{\varepsilon - eV_2}{2T}+ \theta_b
\tanh \dfrac{\varepsilon + eV_2}{2T} \Biggr),
\label{grf}
\end{multline}
where we defined $\Omega \equiv \Omega^R$,
$ P(R_1, R_2)=
|1-R_1 R_2 a^2 - Q[\varepsilon (1+R_1 R_2 a^2) + \Delta a(R_1 +
R_2)]|^2,
$
$Q=\Omega^{-1}\tanh iL\Omega/v_F$, $a=(\Omega-\varepsilon)
/\Delta $, $\theta_b$ and $\theta_c$ equal to unity for
trajectories of respectively Fig. 3b and 3c and to zero otherwise.
As expected, Eq. (\ref{grf}) identifies four different
contributions entering with the corresponding amplitudes and
reflection coefficients. Note that only one out of these
contributions survives in the case of reflectionless interfaces.
In contrast, for weakly transmitting barriers ($R_{1,2} \to 1$)
and $\varepsilon < \Delta$ all four terms enter with equal
prefactors.

As for the function $\tilde g^K$, at SN$_1$ interface it
does not depend on $V_2$ for positive $p_{x_1}>0$.
The values of $g^K$ and $\tilde g^K$ for negative $p_{x_1}<0$
are easily recovered by means of the relation
$g^K(-\bm{p}_F, -\varepsilon, \bm{r},t)=\tilde  g^K(\bm{p}_F, \varepsilon, \bm{r},t)$.

{\it Non-local conductance}. Substituting the results (\ref{gN1}),
(\ref{grf}) into Eq. \eqref{current} we obtain
\begin{gather}
I_1= I_{11}(V_1)+I_{12}(V_2),
\\
I_2= I_{21}(V_1)+I_{22}(V_2).
\end{gather}
Here $I_{11}$ and $I_{22}$ consist of the standard BTK currents
\cite{BTK,Zaitsev} and CAR terms to be specified later and
\begin{multline}
I_{12}(V)=I_{21}(V)=-\frac{G_{N_{12}}}{2e}\int d \varepsilon
\left[\tanh\dfrac{\varepsilon+eV}{2T} -
\tanh\dfrac{\varepsilon}{2T} \right] \\\times (1-{\cal R}_1 |a|^2
)(1-{\cal R}_2 |a|^2 ) \dfrac{1-\tanh^2iL\Omega/v_F}{P({\cal R}_1,
{\cal R}_2)}, \label{12}
\end{multline}
where ${\cal D}_{1,2}\equiv 1-{\cal R}_{1,2}=D_{1,2}(p_F\gamma
_{1,2})$ and $p_F\gamma _{1(2)}$ is normal to the first (second)
interface component of the Fermi momentum for electrons
propagating straight between the interfaces,
\begin{equation}
G_{N_{12}}=\frac{8\gamma_1 \gamma_2{\cal N}_1{\cal N}_2{\cal D}_1{\cal
    D}_2}{R_qp_F^2L^2}
\label{GN12}
\end{equation}
is the non-local conductance in the normal state, ${\cal
N}_{1,2}=p_F^2{\cal A}_{1,2}/4\pi$ define the number of conducting
channels of the corresponding interface, $R_q=2\pi/e^2$ is the
quantum resistance unit. Eq. (\ref{12}) represents the central
result of our paper. This expression fully determines non-local
conductances of our NSN device at arbitrary transmissions of SN
interfaces.

\begin{figure}
\includegraphics[width=7.cm]{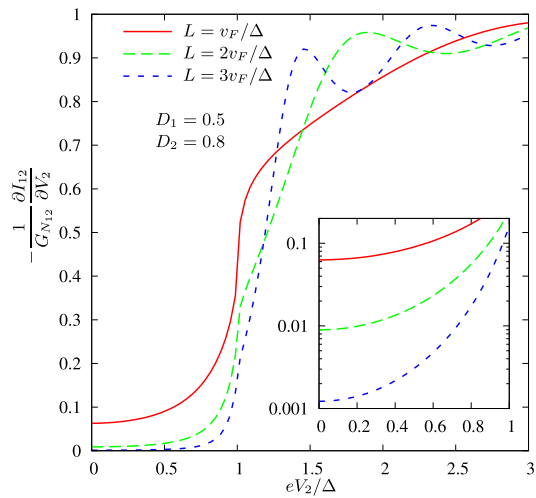}
\caption{(Color online) Differential non-local conductance at $T=0$ as a
function of voltage for $D_1=0.5$, $D_2=0.8$ and different $L$. 
Inset: the same for $eV < \Delta$.}
\end{figure}

The differential non-local conductance evaluated with the aid of
Eq. (\ref{12}) at $T=0$ is presented in Fig. 4 at sufficiently
high interface transmissions. We observe that this quantity
increases sharply around $eV \sim \Delta$ and approaches the
$L$-independent (normal) limit at $eV \gg \Delta$. In the limit
$T, V_{1,2}\ll \Delta$ only subgap quasiparticles contribute and
the differential conductance becomes voltage-independent. We have
$I_{12}=-G_{12}V_{2}$, where
\begin{equation}
\frac{G_{12}}{G_{N_{12}}}= \dfrac{{\cal D}_1 {\cal D}_2
(1-\tanh^2 L \Delta/v_F ) }{
[1+ {\cal R}_1 {\cal R}_2 + ({\cal R}_1 + {\cal R}_2)\tanh L \Delta /v_F]^2}.
\label{nonlcon}
\end{equation}
The value $G_{12}$ (\ref{nonlcon}) gets strongly suppressed with
decreasing ${\cal D}_{1,2}$ and increasing $L$, as also seen in
Fig. 4. Note, that the dependence of $G_{12}$ on $L$ reduces to
purely exponential at all $L$ only in the lowest nonvanishing
order in the transmission of at least one of the barriers, e.g.,
$G_{12} \propto {\cal D}_1^2{\cal D}_2^2 \exp (-2L\Delta /v_F)$
for ${\cal D}_{1,2} \ll 1$, whereas in general this dependence is
slower than exponential at smaller $L$ and approaches the latter
only at large $L \gg v_F/\Delta$.

For a given $L$ the non-local conductance reaches its maximum in
the case of reflectionless interfaces $D_{1,2}=1$. Interestingly,
in this case for small $L \ll v_F/\Delta$ the conductance $G_{12}$
identically coincides with its normal state value $G_{N_{12}}$ at
any temperature and voltage. This result can easily be understood
bearing in mind that for $D_{1,2}=1$ only trajectories indicated
by horizontal lines in Fig. 3b,c contribute to $G_{12}$. For $L\to
0$ there is ``no space'' for CAR to develop on these trajectories
and, hence, CAR contribution to $G_{12}$ vanishes, whereas direct
transfer of electrons between N$_1$ and N$_2$ remains unaffected
by superconductivity in this limit.

The situation changes provided at least one of the transmissions
is smaller than one. In this case scattering at SN interfaces
mixes up trajectories connecting N$_1$ and N$_2$ terminals with
ones going deep into and coming from the superconductor. As a
result, CAR contribution to $G_{12}$ does not vanish even in the
limit $L \to 0$ and $G_{12}$ turns out to be smaller than
$G_{N_{12}}$.

Finally, we would like to briefly address the non-local correction
to $G_{11}$ which arises from the CAR process described by the
term $g_{1,b+c}^K(V_1)$ in Eq. (\ref{gN1}). At $T, V_{1,2}\ll
\Delta$ we have $I_{11}=G_{11}V_{1}$, where
$
G_{11}=G_{1}^{BTK}+\delta G_{11}.
$
Here $G_{1}^{BTK}$ is the standard BTK term
\begin{equation}
G_{1}^{BTK}=\frac{8 {\cal N}_1}{R_q} \left< \frac{|v_{x_1}|}{v_F}
\dfrac{D^2_1(p_{x_1})}{[1+R_1(p_{x_1})]^2} \right>,
\label{BTKcond}
\end{equation}
and for the non-local term we obtain
\begin{multline}
\dfrac{\delta G_{11}}{G_{N_{12}}}=
\dfrac{2(1+ {\cal R}_2) (1-\tanh^2 L \Delta/v_F ) }{
[1+ {\cal R}_1 {\cal R}_2 + ({\cal R}_1 + {\cal R}_2)\tanh L \Delta /v_F]^2}
\\+
\dfrac{ {\cal D}_1
\left[
(1+{\cal R}_2\tanh L \Delta/v_F)^2+
3({\cal R}_2+\tanh L \Delta/v_F)^2
\right]
}{{\cal D}_2
[1+ {\cal R}_1 {\cal R}_2 + ({\cal R}_1 + {\cal R}_2)\tanh L \Delta /v_F]^2}.
\label{CARcorr}
\end{multline}
As compared to the BTK conductance (\ref{BTKcond}) the CAR
correction (\ref{CARcorr}) contains an extra small factor ${\cal
A}_2/L^2$ and, hence, in many cases can be neglected. On the other
hand, since CAR involves tunneling of {\it one} electron through
each interface, for small $D_1 \ll 1$ and $D_2 \approx 1$ we have
$\delta G_{11} \propto {\cal D}_1$, i.e. for $D_1 < ({\cal
A}_2/L^2)\exp(-2L\Delta/v_F)$ the CAR contribution (\ref{CARcorr})
may well exceed the BTK term $G_{1}^{BTK} \propto D_1^2$.

In summary, we have developed a theory of non-local electron
transport in ballistic NSN structures with arbitrary interface
transmissions. Non-trivial interplay between normal scattering,
local and non-local Andreev reflection at SN interfaces yields a
number of interesting properties of non-local conductance which
can be tested in future experiments.

We would like to thank V. Chandrasekhar for communicating the
results \cite{Venkat} to us prior to publication and to A.A.
Golubov for useful discussions at an early stage of this work.



\begin{thebibliography}{99}
\bibitem{FN} The term ``electron cotunneling''
remains relevant only in the tunneling limit loosing its meaning
as soon as one goes beyond perturbation theory and includes
processes to all orders in barrier transmissions.
\end{thebibliography}
\end{document}